\documentclass{ifacconf}

\usepackage{graphicx}      % include this line if your document contains figures
\usepackage{natbib}        % required for bibliography
\usepackage[dvips]{epsfig}    % or this line, depending on which
\usepackage{xcolor}

\usepackage{amssymb}
\usepackage{amsmath}          % Include this line if your
\usepackage{enumerate}

\begin{document}
\begin{frontmatter}

\title{Inversion-free feedforward and feedback control of MSM based actuator with large non-smooth input hysteresis}

\thanks[]{\textcolor[rgb]{0.00,0.00,1.00}{Authors accepted manuscript}}

\author[1]{Michael Ruderman},
\author[2]{Gianluca Giostra},
\author[2]{Matteo Sette}

\address[1]{University of Agder, Department of Engineering Sciences \\ 4879 Grimstad, Norway  \\
Email (correspondence): \tt\small michael.ruderman@uia.no}
\address[2]{University of Padova, Department of Management and Engineering \\ 36100 Vicenza, Italy \\
Email: \tt\small gianluca.giostra@studenti.unipd.it,
matteo.sette.1@studenti.unipd.it}

%%%%%%%%%%%%%%%%%%%%%%%%%%%%%%%%%%%%%%%%%%%%%%%%%%%%%%%%%%%%%%%%%%%%%%%%%%%%%%%%
\begin{abstract}
Dynamic systems with a large and non-smooth hysteresis in the
feedforward channel challenge the design of feedback control since the
instantaneous input gain is varying during the operation, in the worst
case between zero and infinity. Magnetic shape memory (MSM)
actuators with multi-stable transitions represent such untypical
system plant with only the output displacement being measured. This paper
provides a case study of designing the feedforward and feedback
control system for an MSM-based actuator setup with a fairly high
level of the output sensing noise. First, the recently introduced
inversion-free feedforward hysteresis compensator is adapted for
the Krasnoselskii-Pokrovskii operator model. Then, a robust
feedback proportional-integral (PI) loop shaping is performed,
while taking into account the lagging behavior of the low-pass
filtering and system uncertainties. Experimental results show that
the parallel action of feedforward and feedback parts improves
the overall performance of position control.
\end{abstract}


\end{frontmatter}

%===============================================================================

%%%%%%%%%%%%%%%%%%%%%%%%%%%%%%%%%%%%%%%%%%%%%%%%%%%%%%%%%%%%%%%%%%%%%%%%%%%%%%%%
\section{Introduction}
\label{sec:1}

Two-degree-of-freedom (2DOF) control with feedforward may be
required in various actuated systems, especially where the modeled
but not measurable nonlinearities cannot be compensated
effectively by a feedback action only. An example of such
single-input-single-output system is a magnetic shape memory (MSM)
actuator, which is based on active materials with magnetic
field-induced elongation and mechanical stress-driven retraction,
see \cite{ullakko1996}. MSM actuators evoked large attention in
research communities since some two decades, leading to a variety
of the proposed approaches to dynamics modeling, see e.g.
\cite{sarawate2008}, and control, see e.g.
\cite{riccardi2013,ruderman2013}.

Over the decades, enthusiasm for innovation surrounding the MSM
actuator technologies subsided due to the findings regarding their
reversibility, energy efficiency, and uncertainties in the dynamic
and static response. One of the biggest challenges identified in
the conducted research, is a strongly pronounced nonlinear and
multi-stable behavior of the MSM drive elements, that makes their
accurate control a non-trivial task. Nevertheless, the latest
detailed modeling studies, e.g. \cite{ehle2023}, show a continued
interest in the MSM-based actuation technologies, also with regard to
different (alternative) MSM actuator topologies and structures,
like for example multi-stable push-pull linear actuator, e.g.
\cite{courant2024}.

One of the most pronounced characteristic input-output properties
of MSM actuators, is a large and less uniform counterclockwise
input hysteresis between the exciting magnetic field and the
output strain (elongation), while the latter is driven back by an
applied compressive mechanical stress, cf.
\cite{sarawate2008,ruderman2011system}. The \emph{hysteresis}, see
e.g. \cite{Maye03} for basics, is understood as a nonlinear
multi-valued but rate-independent mapping that features a nonlocal
memory. Various nonlinear operator-based hysteresis models are
known (see e.g. \cite{visintin1994} for details) that describe
nonlinear phenomena with memory-affected transitions, while the
so-called Preisach type hysteresis model can be seen as a certain
over-class for different operator-based approaches. For a
recent review on differential equations with hysteresis operators,
existence of their solutions, and stability we also refer to
\cite{leonov2017}.

To control the hysteresis in systems through an inverse
compensation, various approaches (some of which are approximative
and ad-hoc) were proposed over the decades. An overview which
includes iterative hysteresis inversion for classes of Preisach,
Prandtl-Ishlinskii, and Krasnoselskii-Pokrovskii operator models
can be found e.g. in \cite{iyer2009}. Since the Preisach-type
hysteresis models constitute a continuum (or a large number in a
discretized setting) of the individual elementary hysteresis
operators, the question of a practical implementation has always
played one of the key roles for model-based hysteresis control.
Examples of an efficient real-time compatible implementation of
the Preisach hysteresis model are available, e.g.
\cite{ruderman2015}. Also the robust online recursive estimation
methods (\cite{ruderman2018}) were proposed for identifying the
otherwise unknown distribution of the elementary hysteresis
operators, building up the overall input-output hysteresis map.

In this paper, we address the problem of an output displacement
control of the MSM-based actuator (\cite{ruderman2011system}) by
developing a 2DOF control structure whose model-based
feed-forwarding is based on the inversion-free hysteresis
compensation proposed recently in \cite{ruderman2023inversion}.
The targeted MSM actuator system features a fairly high level of
sensing noise (in addition to model uncertainties) that challenges
significantly a feedback control part and makes an hysteresis
compensating feed-forwarding indispensable. The contribution of
the paper is twofold. First, we adapt and prove the inversion-free
feedforward hysteresis compensation from
\cite{ruderman2023inversion} to the Krasnoselskii-Pokrovskii
hysteresis model with non-smooth transitions and loss of strict 
monotonicity. Second, we demonstrate a practical experimental case
study which highlights several hurdles of the essential model
identification, design of the output feedback control part, and
realization of the 2DOF control architecture.

%%%%%%%%%%%%%%%%%%%%%%%%%%%%%%%%%%%%%%%%%%%%%%%%%%%%%%%%%%%%%%%%%%%%%%%%%%%%%%%%
\section{Inversion-free feedforward hysteresis compensation}
\label{sec:2}

\subsection{Internal model based approximation of inverting}
\label{sec:2:sub:1}

\begin{figure}[!h]
\centering
\includegraphics[width=0.7\columnwidth]{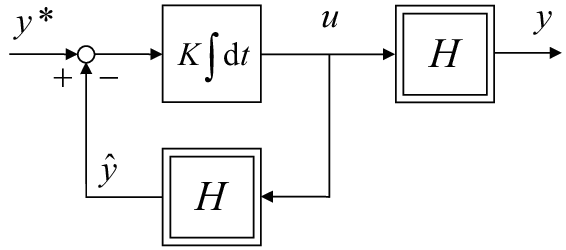}
\caption{Inversion-free feedforward hysteresis control.}
\label{fig:invfreecomp}
\end{figure}

Following the internal model principle (IMP), see
\cite{francis1976}, that requires incorporating in the feedback
path a suitably reduplicated model of the dynamic structure of
disturbance, an approximation of inverting the hysteresis map $H$
was proposed in \cite{ruderman2023inversion} as it is shown in Fig. \ref{fig:invfreecomp}.
Recall that attractive properties of IMP carry over also to
general nonlinear case, see \cite{economou1986}. In
\cite{ruderman2023inversion}, it was shown that since a hysteresis
map $H$ does not have any internal state dynamics and, thus, can
be seen at each input-output operation point as a linear
combination of the known bounded bias $Y_0$ and finite gain
$\gamma = \partial y / \partial u$, the $\hat{y}(t) \rightarrow
y^{\ast}(t)$ and thus $u(t)$ approximates an inverse of the
hysteresis map $H: \, u \mapsto y$. It was demonstrated in
\cite{ruderman2023inversion} that the integrator gain $K$ can be
assigned arbitrary large, theoretically, and the loop error
magnitude $|y^{\ast} - \hat{y}|$ goes to 0 at steady-state, and
becomes 1 at higher frequencies of $y^{\ast}(t)$, provided $\gamma
> 0$ everywhere.

\subsection{Krasnoselskii-Pokrovskii hysteresis model}
\label{sec:2:sub:2}

\begin{figure}[!h]
\centering
\includegraphics[width=0.5\columnwidth]{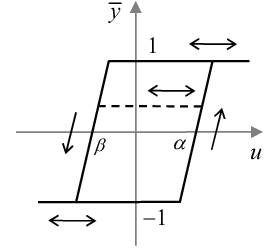}
\caption{Krasnoselskii-Pokrovskii hysteresis operator.}
\label{fig:KPoper}
\end{figure}

\begin{figure}[!h]
\centering
\includegraphics[width=0.98\columnwidth]{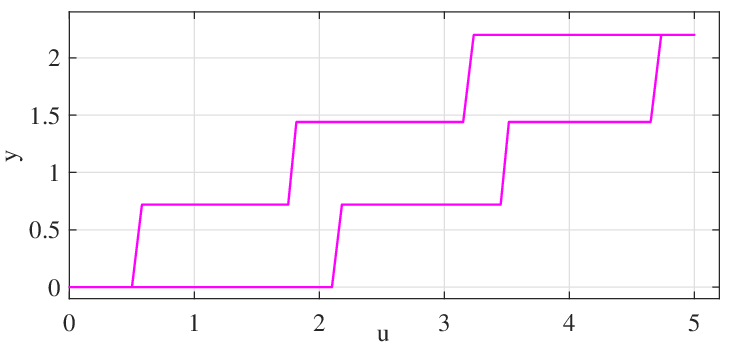}
\caption{Input-output hysteresis by KP model with $N=3$.}
\label{fig:hysKP}
\end{figure}

The so called Krasnoselskii-Pokrovskii (KP) hysteresis operator,
see \cite{krasnosel1989} for basics, is a memory-affected
Preisach-type operator (cf. \cite{Maye03}) whose input-output
transitions are as depicted in Fig. \ref{fig:KPoper}.
The nonlinear static map is given by
\begin{equation}
\bar{y} = h[u,y_0, \alpha, \beta], \label{eq:2:2:1}
\end{equation}
where the initial state is captured by $y_0$ and the spatial
parameters are the output zero-crossing values $\alpha > \beta$.
Note that if $\alpha = \beta$, the the hysteresis operator
\eqref{eq:2:2:1} reduces to the saturated linear unity slope. Also
we note that this zero-crossing-type parametrization only follows
some conventions of the hysteresis literature (cf. e.g.
\cite{visintin1994}, \cite{Maye03}), while $\alpha, \beta$
parameters can also be assigned for instance to the threshold
values where the slope transitions (see Fig. \ref{fig:KPoper})
begin, cf. e.g. \cite{iyer2009}. The overall KP hysteresis model
is obtained by a weighted continuum of the elementary operators
\eqref{eq:2:2:1}, in accord with the Preisach-type hysteresis
model principles (\cite{visintin1994,Maye03}), as
\begin{equation}
y(t) = \iint \limits_{\alpha \geq \beta} \rho(\beta,\alpha)
h\bigl[u(t),y_0, \alpha, \beta \bigr] d \beta d \alpha.
\label{eq:2:2:2}
\end{equation}
Here $\rho : P \rightarrow \mathbb{R}_{+}$ is a positive,
integrable function on $P = \bigl\{(\beta,\alpha): \: \alpha \geq
\beta \bigr\}$ which constitutes the density of the elementary
operators distribution over the so called Preisach plane $P$. In
case of a finite number $N$ of the elementary hysteresis operators
\eqref{eq:2:2:1}, the KP hysteresis model is written as a weighted
superposition
\begin{equation}
y(t) = \sum \limits_{n=1}^{N} \rho_n \, h\bigl[u(t),y_0, \alpha_n,
\beta_n \bigr]. \label{eq:2:2:3}
\end{equation}
Further we note that the KP hysteresis operator \eqref{eq:2:2:1}
can equally be implemented as a standard Prandtl-Ishlinskii
play-type operator (\cite{visintin1994}) subject to the input
shift by $\delta$ and output saturation at the magnitude $m$. With
the play-slot width $w$ and the slope gain $\gamma$, it results in
an equivalent memory-affected input-output map
\begin{equation}
\bar{y} = \gamma \, h \bigl[ (u+\delta), y_0, w, m \bigr].
\label{eq:2:2:4}
\end{equation}
One can show that $h$ is transformable back from \eqref{eq:2:2:4}
to \eqref{eq:2:2:1} by $\alpha = - \delta + 0.5 w $ and $\beta = -
\delta - 0.5 w $, while for the KP operator shown in Fig.
\ref{fig:hysKP} the saturation magnitude is $m=0.72$. Below, we take
advantage of the KP representation \eqref{eq:2:2:4}, not only for
the sake of implementation but also for the following proof of
convergence of the feedforward hysteresis control, cf. section
\ref{sec:2:sub:1}. An input-output hysteresis map described by the
KP model with $N = 3$ is exemplary shown in Fig. \ref{fig:hysKP}
in approaching the experimentally identified MSM hysteresis, cf.
with Fig. \ref{fig:exphysteresis}.

\subsection{Feedforward compensation}
\label{sec:2:sub:3}

Below, a model-in-the-loop stability of the inversion-free
feedforward compensator when using the KP-based hysteresis
mapping $H$ (see Fig. \ref{fig:invfreecomp}) will be demonstrated.
In that sense we are to show that $(y^{\ast} - \hat{y})
\rightarrow 0$ so that the dynamic quantity $u(t)$ approximates
the inverse of the KP-based hysteresis mapping $H: \, u \mapsto y$,
cf. Fig. \ref{fig:invfreecomp}.

The feedforward compensator (Fig. \ref{fig:invfreecomp}) implies
\begin{equation}
\dot{u} = \frac{1}{K} \bigl(y^{\ast} - H(u) \bigr).
\label{eq:2:3:1}
\end{equation}
Note that \eqref{eq:2:3:1} belongs to a generic class of the
forced first-order ordinary differential equations coupled with the
hysteresis operator
\begin{equation}
\dot{u} + H(u) = f(t), \label{eq:2:3:2}
\end{equation}
for which the uniqueness of solutions for the Cauchy problem was
shown in \cite{kopfova1999}. The results obtained in
\cite{kopfova1999} are valid for any kind of hysteresis operators
with monotone curves and a continuous $f(t)$ on $[0, T]$. Since
the strict monotonicity is violated for the KP hysteresis model with a
finite number of elementary operators, cf. section
\ref{sec:2:sub:2}, we are to analyze \eqref{eq:2:3:1} in more
detail. For the sake of analysis and without loss of generality,
the differential equation \eqref{eq:2:3:1} can be piecewise
reproduced by
\begin{equation}
\dot{u} = \frac{1}{K} \bigl(y^{\ast} - (Y_0 + \gamma \, u) \bigr).
\label{eq:2:3:3}
\end{equation}
Here $Y_0$ and $\gamma$ are the instantaneous output bias and
slope gain of the hysteresis map.

Assume first, the hysteresis state of the KP model is within a
\emph{play-slot}. This implies $\gamma = 0$ and converts \eqref{eq:2:3:3}
into
\begin{equation}
\dot{u} = \frac{1}{K} \bigl(y^{\ast} - Y_0 \bigr).
\label{eq:2:3:4}
\end{equation}
It is evident that \eqref{eq:2:3:4} always drives $u(t)$ out from
the play-slot and, thus, brings KP hysteresis model state to the \emph{slope},
with an instantaneous $\gamma \neq 0$, cf. Fig.
\ref{fig:KPoper}.

Afterwards, consider the KP hysteresis model $H$ on the slope
(i.e. $\gamma \neq 0$) and assume, for instance and without loss
of generality, $(y^{\ast} - Y_0 \bigr) > 0$. Then,
\eqref{eq:2:3:3} is rewritten as
\begin{equation}
K \dot{u} + \gamma u  = y^{\ast} - Y_0. \label{eq:2:3:5}
\end{equation}
Two facts can be recognized for the homogenous (i.e.
left-hand-side) part of the differential equation
\eqref{eq:2:3:5}: (i) it is asymptotically stable for any $K,
\gamma > 0$ that is always true, (ii) it provides a monotonic
exponential transient (i.e. without oscillatory overshoots) as
long as $(y^{\ast} - Y_0) \neq 0$. Then, two scenarios can appear
for a continuously increasing $u(t)$ and depending on the
instantaneous $y^{\ast}$ value. Either the KP hysteresis model
reaches again a play-slot (not unusual for $N > 1$), or $Y_0 \neq
y^{\ast}$ will increase $u(t)$ asymptotically towards an
equilibrium
\begin{equation}
u_0 = \frac{y^{\ast} - Y_0}{\gamma}, \label{eq:2:3:6}
\end{equation}
for which one can show (geometrically) that $\gamma u + Y_0 =
\hat{y} \rightarrow y^{\ast}$. This provides an asymptotic
approximation of $u(t)$ to the inverse of the KP hysteresis map.

\begin{figure}[!h]
\centering
\includegraphics[width=0.98\columnwidth]{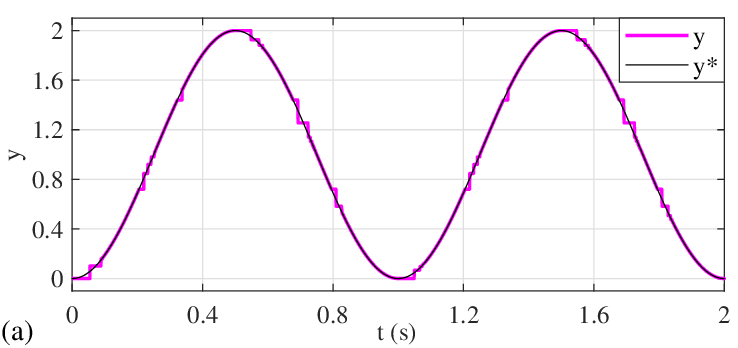}
\includegraphics[width=0.98\columnwidth]{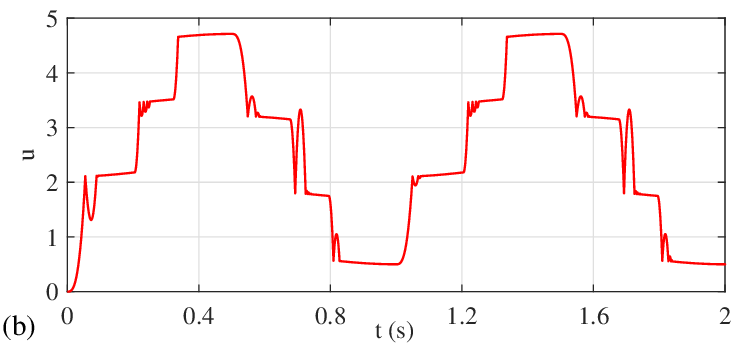}
\caption{Feedforward compensated hysteresis, (a) hysteresis
output versus reference, (b) compensator output.}
\label{fig:comphyst}
\end{figure}

A numerical example of the inversion-free feedforward hysteresis
control with $K=2000$ is demonstrated in Fig.
\ref{fig:comphyst}, for the KP hysteresis model with $N=3$ as in
Fig. \ref{fig:hysKP}. The output $y$ versus the reference
$y^{\ast}$ are shown in (a), while the inverse compensator output
$u(t)$ is shown in (b).

%%%%%%%%%%%%%%%%%%%%%%%%%%%%%%%%%%%%%%%%%%%%%%%%%%%%%%%%%%%%%%%%%%%%%%%%%%%%%%%%
\section{MSM actuator system}
\label{sec:3}

\subsection{Mechatronic setup of MSM actuator}
\label{sec:3:sub:1}

\begin{figure}[!h]
\centering
\includegraphics[width=0.3\columnwidth]{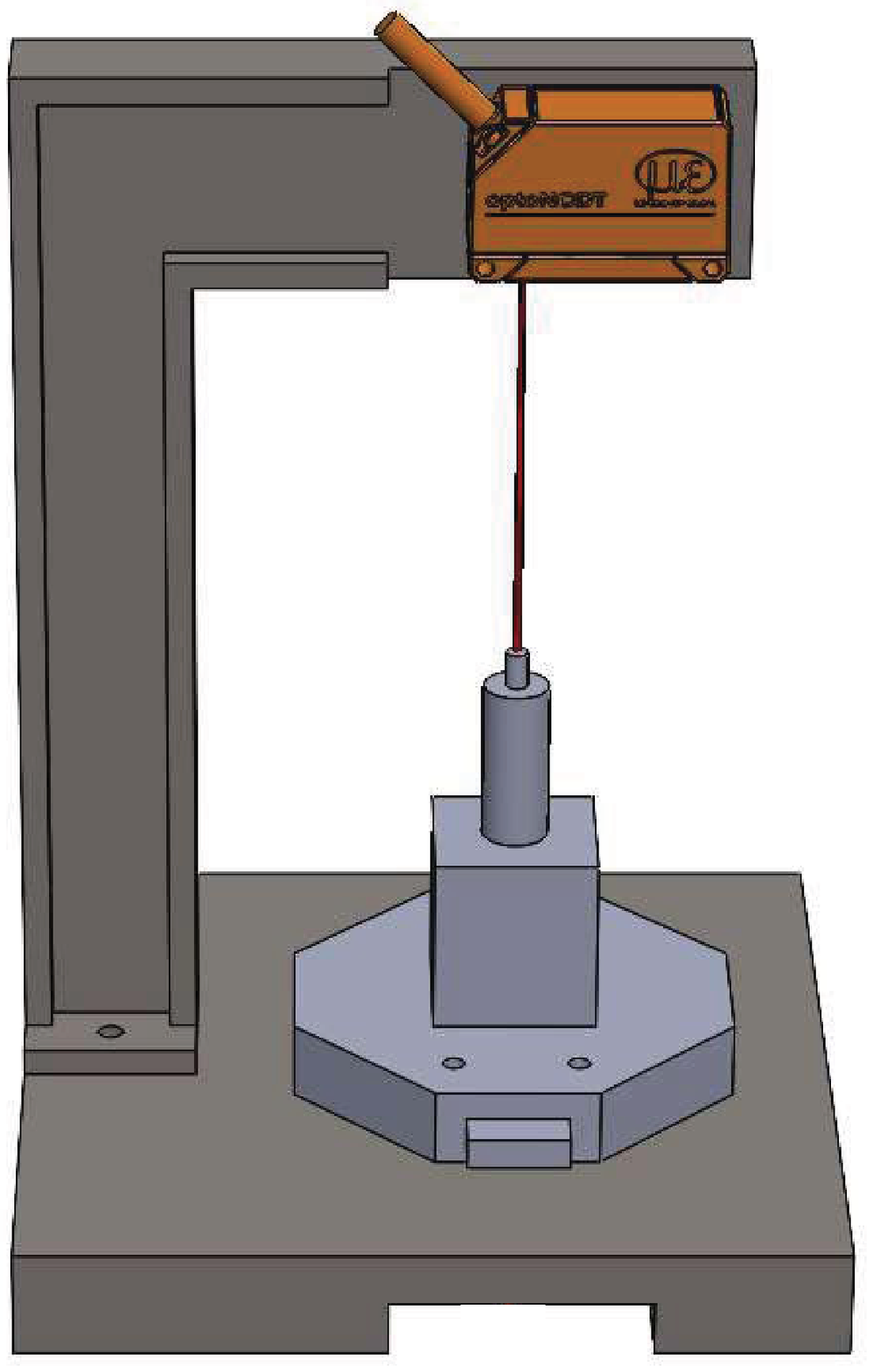} \hspace{1mm}
\includegraphics[width=0.66\columnwidth]{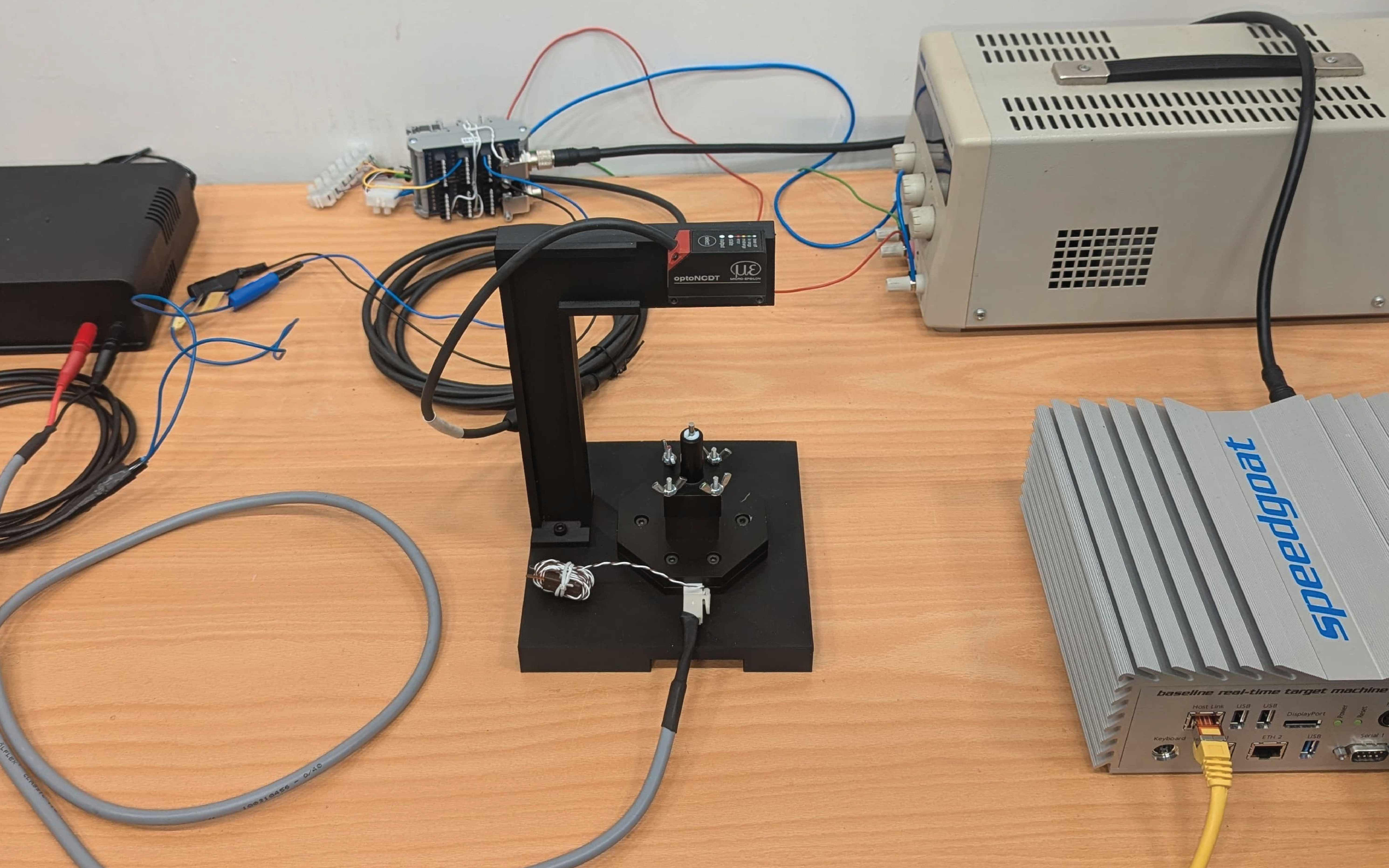}
\caption{MSM actuator system setup: excerpt from CAD drawing
(left) and laboratory view (right).} \label{fig:msmactuator}
\end{figure}

The MSM actuator setup used in this study is shown in Fig.
\ref{fig:msmactuator}, where an excerpt from the CAD drawing of
assembly is depicted on the left and the laboratory view is
illustrated on the right. The latter includes a low-level
(electric circuit-based) current control unit, a mechanical mount
with the laser sensor and actuator itself, wiring and connectors,
a DC power supply, and the SpeedGoat realtime board connected to a
host computer. The set sampling frequency is 2 kHz. The nominal
repeatability of the laser sensor is 8 $\mu$m, while the maximal
MSM stroke is about 500 $\mu$m.
The realtime hardware board provides the analog command $i^{\ast}
\in [0,\ldots,5]$ A to the current controller, while the unknown
time constant of the current feedback loop is below 1 ms. The MSM
actuator, produced by AdaptaMat Ltd company (see
\cite{ruderman2013}), integrates a single Ni-Mn-Ga MSM element of
the size $20 \times 2.5 \times 1.0$ mm$^3$. The nominal actuator
characteristics from the manufacturer's information sheet are the
output force 2.5--3.5 N, the field strength required for maximum
strain about 400--500 kA/m, the operating frequency up to 1--2 kHz,
the operating temperature of maximal 40 C, and the typical stroke
of 3--5 \%. The mechanical screw of the pre-stressing spring of the
MSM actuator was adjusted in a sequence of experiments so as to
provide the maximal possible strain (elongation) of the MSM
element with the reproducible returning (compression) cycles, cf.
Fig. \ref{fig:exphysteresis}.

\subsection{Input-output system modeling}
\label{sec:3:sub:2}

A control-oriented input-output system modeling takes into account
both, the second-order linear dynamics of mechanical part of
the actuator, which represents approximately a mass-spring-damper
system, and the hysteretic MSM transducer. The simplified MSM
actuator model is given by (\cite{ruderman2011system}):
\begin{equation}
m \ddot{x}(t) + d \dot{x}(t) + k x(t) = \kappa \, H\bigl[i(t),
\sigma(t) \bigr], \label{eq:3:1}
\end{equation}
where the excitation current $i$ in the electro-magnetic coils
serves as a controllable system input. The mechanical parameters
of the lumped mass, damping, and restoring spring $m,d,k > 0$, respectively, 
are assumed to be the known constants. The overall input gain is
$\kappa > 0$, and the counteracting internal (respectively
external) mechanical stress is denoted by $\sigma$. Worth noting
is that $\sigma$ is neither measurable nor observable and, thus,
its deterministic effect on the MSM hysteresis behavior is
excluded from the modeling and is, therefore, considered as a
bounded disturbance factor. The static MSM hysteresis map $H: i
\mapsto y \,$ is a multi-valued (i.e. memory-affected) operator
described by the KP model, cf. section \ref{sec:2:sub:2}. Note
that for the sake of simplicity, the overall input-output steady-state gaining factor can be accommodated in one and the same constant $\tilde{\kappa}$. Following to that, the $x = \tilde{\kappa} H(i)$ hysteresis map can be identified under quasi-static operation
conditions.

\subsection{Identified system behavior}
\label{sec:3:sub:3}

\begin{figure}[!h]
\centering
\includegraphics[width=0.98\columnwidth]{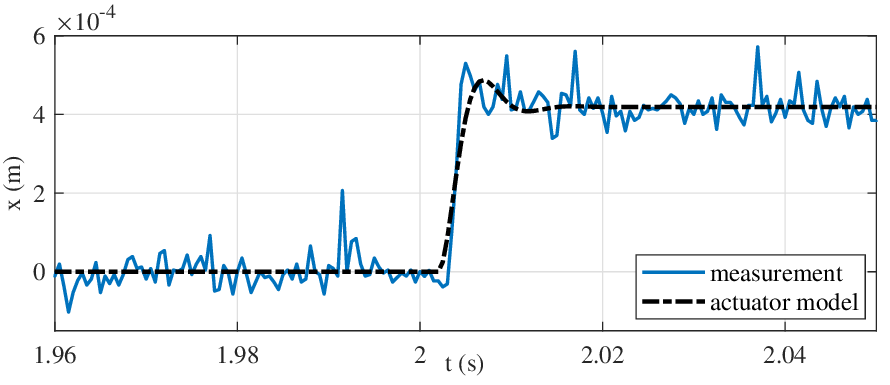}
\caption{Actuator step response: measured versus modeled.}
\label{fig:stepresponse}
\end{figure}

The measured step response of the MSM actuator is shown in Fig.
\ref{fig:stepresponse}. One can recognize a relatively high level
of the overlapping sensor and process noise. Therefore, the
identification of linear system parameters, cf. \eqref{eq:3:1},
requires an averaging broadband excitation and measurement of the
frequency response function (FRF) characteristics.

\begin{figure}[!h]
\centering
\includegraphics[width=0.98\columnwidth]{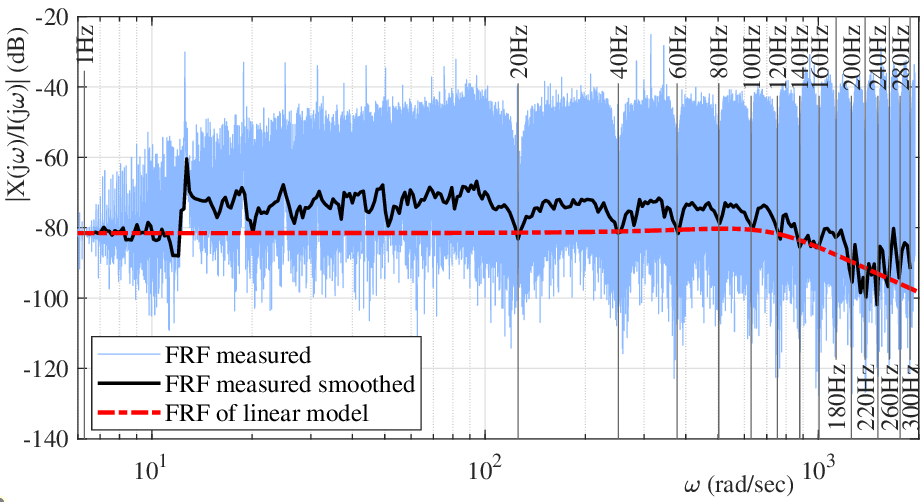}
\caption{Frequency response function (FRF) of MSM actuator:
measured, smoothed, and linear model fitted.}
\label{fig:frfmeasured}
\end{figure}

The sequence of the sinusoidal inputs of different frequencies
between 1 and 300 Hz, each one revealing multiple steady-state
periods, was used for exciting the MSM actuator. The resulted FRF,
determined from the input and measured output data, is shown in
Fig. \ref{fig:frfmeasured}. The smoothed FRF estimate with a
variable frequency resolution using the spectral analysis (i.e. using
\emph{spafdr} algorithm of Matlab system identification toolbox)
is shown over the raw FRF data. The linear transfer function model
\begin{equation}
G(s) \equiv \frac{x(s)}{i^{\ast}(s)} = \frac{45.57 \exp(-0.002
s)}{s^2 + 737.9s + 5.439 \times 10^5}, \label{eq:3:2}
\end{equation}
which is least squares identified on the collected broadband
experimental data, is also shown by amplitude response in Fig.
\ref{fig:frfmeasured}. Note that the command current $i^{\ast}$ is
the available input quantity, cf. \eqref{eq:3:1}, while an
additional time-delay of 0.002 sec was also identified. The latter
can be seen as a summarized time delay of the embedded low-level
current control and all analog-digital and digital-analog
converters (including the displacement sensor) in the loop. The
step response of the identified model \eqref{eq:3:2} is also
plotted over the measurement in Fig. \ref{fig:stepresponse} for
the sake of comparison. The second-order dynamics in
\eqref{eq:3:2} highlights the natural frequency $\omega_n = 737$
rad/sec and damping ratio $\zeta = 0.5$.

\begin{figure}[!h]
\centering
\includegraphics[width=0.98\columnwidth]{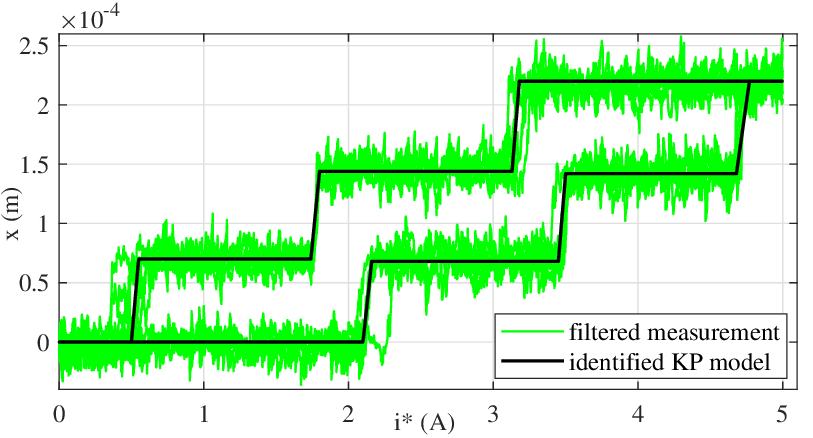}
\caption{Measured and modeled input-output hysteresis.}
\label{fig:exphysteresis}
\end{figure}

The measured MSM hysteresis, see Fig. \ref{fig:exphysteresis}, is
obtained by applying a triangular wave with amplitude 5 A and
frequency 0.1 Hz. The signal was designed to provide a possibly
uniform and quasi-static system response during both the rising
and falling phases. The measured output data was low-pass filtered
with the cutoff frequency of 10 Hz. With the structural assumption
of $N=3$, the threshold value, slope gain, and saturation
magnitude parameters were identified by using the measured
input-output data. The KP modeled hysteresis response $x =
\tilde{\kappa} H(i^{\ast})$ is shown in Fig.
\ref{fig:exphysteresis} over the measurements.

\begin{figure}[!h]
\centering
\includegraphics[width=0.98\columnwidth]{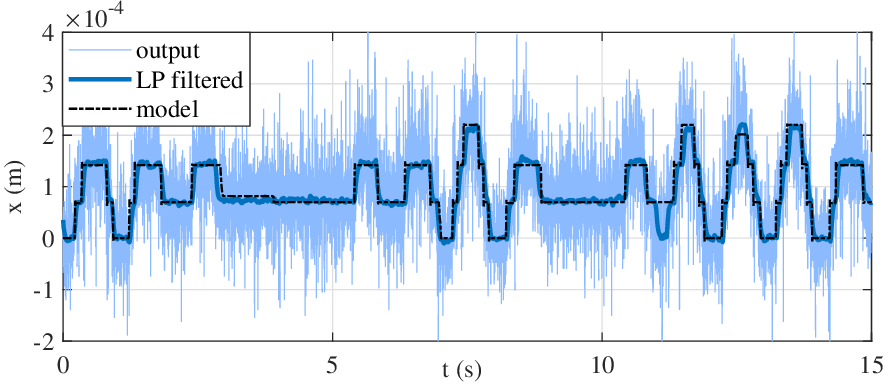}
\caption{Model response of MSM actuator versus the measured and
low-pass filtered output time series.} \label{fig:modelresponse}
\end{figure}

The overall identified input-output model is a serial connection
of the linear transfer function and the unity-saturated input KP
hysteresis model (that means without $\tilde{\kappa}$), yielding
$G(s)\mathcal{L} \bigl[ H(i^{\ast}) \bigr]$ where $\mathcal{L}$ is
Laplace transform of the input signal. The model response is
compared with the measured output in Fig. \ref{fig:modelresponse}.
The latter is obtained as response to a random-amplitude input
excitation, with 1 Hz carrier frequency. Both, the raw measured
data and the same but low-pass filtered data with 10 Hz cutoff
frequency are plotted versus the identified model response.

%%%%%%%%%%%%%%%%%%%%%%%%%%%%%%%%%%%%%%%%%%%%%%%%%%%%%%%%%%%%%%%%%%%%%%%%%%%%%%%%
\section{Experimental control evaluation}
\label{sec:4}

\subsection{Control system with two degrees of freedom}
\label{sec:4:sub:1}

\begin{figure}[!h]
\centering
\includegraphics[width=0.9\columnwidth]{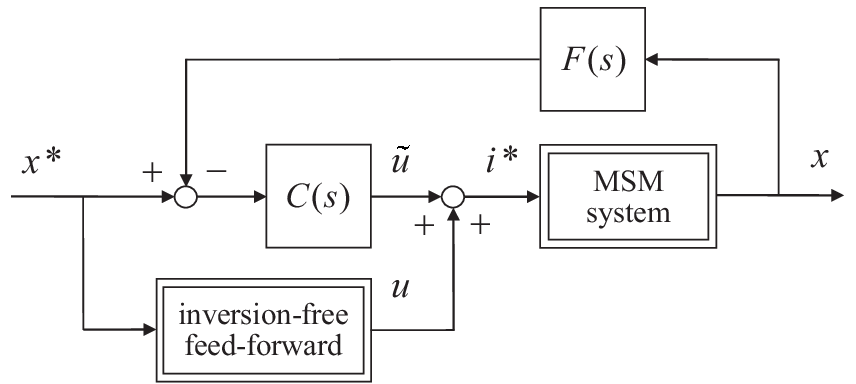}
\caption{2DOF control system of MSM actuator.}
\label{fig:controlstruct}
\end{figure}

For an MSM actuator system, where the input nonlinearity is
substantial and memory-dependent over the entire input-output
operating range, a model-based feed-forwarding can be used to
bring the system close to an operating point without dynamics of the
feedback loop. This enables developing a control system with two
independent degrees of freedom (2DOF), i.e. feedforward and
feedback in their classical sense. Note that a previous comparative
study in \cite{ruderman2013} revealed such control architecture as
most competitive for an MSM-based actuator. Since an inverse
nonlinearity mapping drives an uncertain system only to some
vicinity of an operating point, it is a linear feedback control
part which serves to compensate for the residual control errors. The
resulted 2DOF control scheme is shown in Fig.
\ref{fig:controlstruct}, while an inversion-free feed-forwarding
is realized by an approximative approach described in section
\ref{sec:2}.
The linear feedback controller, designed as described below in section
\ref{sec:4:sub:2}, is denoted by $C(s)$, while $F(s)$ is a
low-pass filter, which is necessary due to a high level of the
displacement sensor noise, cf. section \ref{sec:3:sub:3}. Note
that the drawn MSM system block includes the MSM actuator itself
as well as the contact-less laser sensor and the embedded
current controller which is connected to the excitation coils.

\subsection{Feedback control design}
\label{sec:4:sub:2}

For the linear control subsystem, a standard PI controller
\begin{equation}
C(s) = \frac{\tilde{u}(s)}{e(s)} =  K_p + \frac{K_i}{s}
\label{eq:4:1}
\end{equation}
is assumed, where $K_p, K_i > 0$ are the design parameters and the
feedback control error is $e$, cf. Fig. \ref{fig:controlstruct}.
Note that the second-order system dynamics is assumed as
sufficiently damped, while the integral control action should
allow to compensate (at least partially) for the hysteresis
nonlinearity, cf. e.g. \cite{ruderman2010}. Further we note that a
second-order low-pass filter
\begin{equation}
F(s) = (\mu s + 1)^{-2}  \label{eq:4:2}
\end{equation}
is applied as a necessary one due to a fairly large noise
in the output displacement signal $x(t)$, cf. Figs.
\ref{fig:modelresponse}, \ref{fig:controlstruct}. The filter
cut-off frequency is assigned to 10 Hz. A standard control loop
shaping procedure was performed so that the targeted phase margin
(see e.g. \cite{aastrom2021} for details) of the open-loop
transfer function
\begin{equation}
L(s) \equiv C(s) G(s) F(s)  \label{eq:4:3}
\end{equation}
yields a value $> 60$ deg. That one is assumed to be sufficient for a
robust feedback stabilization. The determined, this way, control
gains are $K_p = 1.13 \times 10^4$, $K_i = 3.06 \times 10^5$.

\subsection{Comparison of position control results}
\label{sec:4:sub:3}

Below, the control response is experimentally evaluated by
comparing the PI feedback control only (designed as in section
\ref{sec:4:sub:2}) and the same PI feedback control combined with
feed-forwarding (cf. section \ref{sec:4:sub:1}). In addition, the
response of the same feedforward control only is also shown in
case of a smooth (sinusoidal) reference signal.

\begin{figure}[!h]
\centering
\includegraphics[width=0.98\columnwidth]{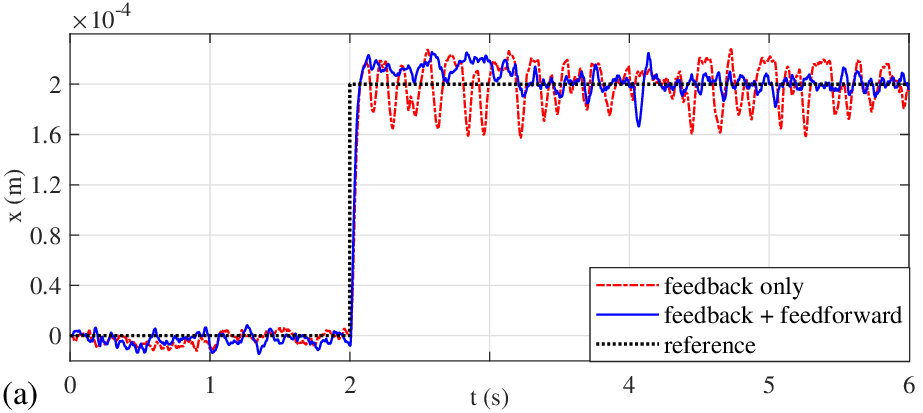}
\includegraphics[width=0.98\columnwidth]{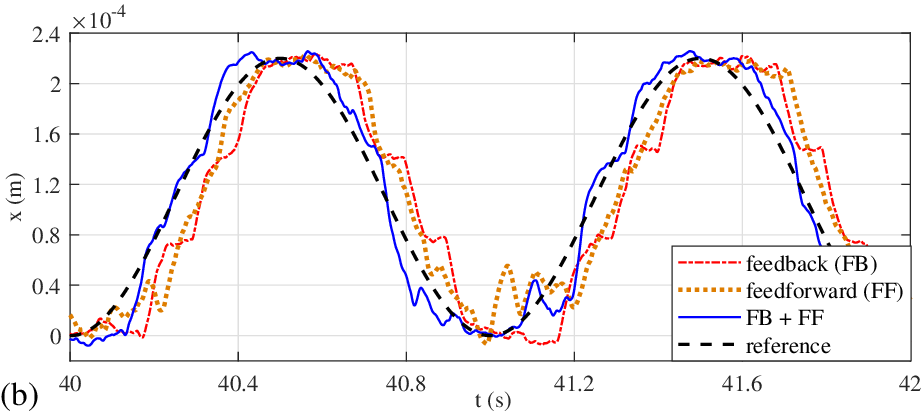}
\caption{Comparison of controlled output: step response (a) and
sinusoidal response of 1 Hz (b).} \label{fig:controlresults}
\end{figure}
First, the measured step response are compared in Fig.
\ref{fig:controlresults} (a). Both, the single PI feedback control
and the 2DOF control system achieve the certain steady-state
accuracy, since employing an integral feedback control action. At
the same time, one can recognize that the single feedback control
provides a much larger fluctuation around the constant reference
value. This is not surprising since the integral action over- and
respectively undershoots every time the reference operation
point due to a large and non-smooth input hysteresis nonlinearity,
cf. Fig. \ref{fig:exphysteresis}.

Next, a sinusoidal reference trajectory (with 1 Hz frequency) is
tracked, see Fig. \ref{fig:controlresults} (b). Here one
can recognize that the PI feedback control and the single
feed-forwarding are inferior comparing to the 2DOF combination of
both. 2DOF control follows closer the reference sinusoidal
during both ascending and descending branches, and has a lower
transient peaking during the direction reversals.

%%%%%%%%%%%%%%%%%%%%%%%%%%%%%%%%%%%%%%%%%%%%%%%%%%%%%%%%%%%%%%%%%%%%%%%%%%%%%%%%
\section{Conclusions}
\label{sec:4}

In this paper, we developed a case study of controlling an
MSM-based actuator by using the two-degree-of-freedom (2DOF) approach
-- with PI output feedback regulator and inversion-free
feedforward compensator of a large input hysteresis. The latter is
adapted from the recently proposed internal model based
compensator (\cite{ruderman2023inversion}) for the
Krasnoselskii-Pokrovskii (KP) hysteresis operator model. A
mechatronic system design and the minimal required input-output
system model identification are shown, while the level of the
sensing and process noise is relatively high in the setup. The
convergence, and thus the model loop stability, of the KP-based
inversion-free feedforward hysteresis control scheme is shown.
Then, the 2DOF control structure is introduced and the feedback
loop is shaped by using the standard stability margin criteria. The
experimental evaluation discloses that the used 2DOF control
scheme is superior comparing to the single PI feedback control or
feedforward compensator only.

%%%%%%%%%%%%%%%%%%%%%%%%%%%%%%%%%%%%%%%%%%%%%%%%%%%%%%%%%%%%%%%%%%%%%%%%%%%%%%%%
\section*{Acknowledgement}
\label{sec:5} This work was partially supported by ERASMUS+
program: Student Mobility for Studies.

\bibliography{references}             % bib file to produce the bibliography
                                                     % with bibtex (preferred)

\end{document}